# Formation of non-cubic nanoparticles from cubic MgO in intensified self-burning of magnesium

Sukbyung Chae[1], Peter V. Pikhitsa[1*], Seungha Shin[1,2], Chang Hyuk Kim[1,3], Sekwon Jung[1,4] and Mansoo Choi[1*]

[1] *Global Frontier Center for Multiscale Energy Systems, Division of WCU Multiscale Mechanical Design, School of Mechanical and Aerospace Engineering, Seoul National University, Seoul 151-744, Korea*

[2] *Department of Mechanical Engineering, The University of Michigan, Ann Arbor, Michigan 48109-2125, United States*

[3] *Department of Mechanical Engineering, The University of Minnesota, Minneapolis, Minnesota 55455, United States*

4 *Automotive Research & Development Division, Hyundai Motor Group, Yongin-si, Gyeonggi-do 446-912, Korea*

**Abstract**

When Mg metal burns in air the resulting rock-salt MgO smoke consists of perfect [100] cubes of about 100 nm. On contrast, we found that intensification of self-burning of Mg micropowder either by injecting it into oxy-hydrogen diffusion flame or under an infrared laser beam switches the growth mechanism producing mostly single-crystalline spheres and terraced nanoparticles. MgO molecule condensation onto primary spherical nanoparticles can account for generation of terraced nanoparticles with regular steps proportional to the nanoparticle size.

*Corresponding authors

E-mail addresses: mchoi@ snu.ac.kr; peter@snu.ac.kr

# 1. Introduction

MgO is one of the most intensively studied materials because of purely ionic nature and simple cubic rock-salt structure. MgO nanoparticles can be used for the construction of functional nanostructures and for the investigation of surface reactivity on oxides. Since equilibrium MgO [100] surfaces are considered chemically inert, the defects (either point defects, such as ion vacancies and coordinatively unsaturated ions in extended defects, such as terrace steps) determine the properties of the surface. Five-, four- and three-coordinated sites, i.e. 5C, 4C and 3C, characteristic for terraces, terrace steps and kinks, respectively, are known to be responsible for electronic states of MgO nanoparticles [1]. One and two electron anion vacancies ($F^+$ and $F^0$ centers, respectively) are also sensitive to morphological features of MgO surface [2].

However, the shape control has been hardly possible for pure MgO nanoparticles: Mg metal burning in air [3,4] or diluted oxygen (with argon or nitrogen) always results in perfectly cubic nanoparticles [5]. High-temperature annealing of hexagonal $Mg(OH)_2$ leads to piles of sintered nanocubes [6]. Prolonged water etching MgO cubic nanoparticles produces [110] facets made exclusively of 1 nm [100] steps [5]. Yet, a slight modification of the growth process – the presence of Si substrate surface where MgO molecules, obtained in the gas reaction $2Mg(vapor)+O_2(gas) \rightarrow 2MgO(vapor)$, can deposit - is known to induce the growth of MgO wires and branches with square cross-sections [7,8]. It indicates that some interference into the reaction and condensation process can be fruitful in changing the morphology of MgO nanoparticles.

Recently, we reported on a novel UV-Vis two-band luminescence of non-cubic (terraced and spherical) MgO nanoparticles originating from a divacany (P-center) [9]. We showed that the P-center could appear only when a growth mechanism of nanoparticles changed by the change of the flame conditions. Such a change was indicated by the appearance of terraced and spherical MgO nanoparticles. Yet, the details of the alternative growth mechanism that lies under the phenomenon were only outlined in Ref. [9] and have not been studied deeply.

In the present work we report on non-cubic (spherical and terraced) MgO nanoparticles (see Fig. 1 and 2) generated under intensification of Mg self-burning either by oxy-hydrogen diffusion flame when Mg fine powder (flakes of size 100 μm) is injected or under a powerful infrared $CO_2$ laser and discuss the growth mechanism leading to such shapes.

## 2. Experimental

The left panel in Fig. 1 shows a part of the flame reactor that we used for flame metal combustion [10]. Fine powder was entrained through the central nozzle into the flame feather either with inert gases (Ar, $N_2$) or with $O_2$. Typical flow rates for oxygen, hydrogen and carrier gas were of order of 1 lpm and the gas velocities of 1 m/sec.

Non-cubic nanoparticles are generated in oxy-hydrogen flame in both cases of inert and oxygen carrier gases. High-resolution transmission electron microscopy (HR-TEM) images of terrace or spherical nanoparticles confirm that the MgO lattice for resultant nanoparticles remains perfectly cubic (Fig. 1 and 3(a)). Both spherical and terraced nanoparticles are single-crystalline nanoparticles with the Mg:O ratio 1:1 according to EDX spectroscopy data (Fig. 3(b)). Yet, only in the case of oxygen carrier gas, after a close inspection of nanoparticles, we could get a clue to the growth mechanism by distinguishing exceedingly small and nearly perfect spheres of MgO that may act as seeds for larger terraced nanoparticles and larger spheres at the later stage of the nanoparticle growth process (Fig. 2). This scenario may be similar to the growth of MgO nanowires from precursor nanocubes that nucleate on a Si substrate [7,8], only now the role of the precursor is played by the incipient spherical nanoparticles.

We performed a control experiment with burning Mg in pure oxygen which proved that Mg self-burning in $O_2$ atmosphere induces exclusive fast growth of mostly defective surfaces and therefore, additionally to cubes, the resulting MgO smoke contains a portion of nearly perfect spheres. Otherwise, we intensified Mg self-burning in air by irradiating a burning Mg pellet with a 2.8 kW cw infrared $CO_2$ laser which now resulted in overwhelming number of spherical nanoparticles.

## 3. Results and discussion

Since long ago it has been known that there are two possible reactions that lead to solid MgO nanoparticles from Mg burning [11]: (I) $MgO_{(g)} \rightarrow MgO_{(s)}$, with enthalpy -566.3 kJ/ mol at 1900 K; (II) $(MgO)_n + Mg_{(g)} + 1/2\ O_2 \rightarrow (MgO)_{n+1}$, with larger enthalpy -729.3 kJ/ mol at 1900 K . It is believed that only the first process, thus only the condensation from supersaturated MgO gas into MgO solid, takes place [11-13]. Growth (I) governed only by condensation of MgO molecules leads to incipient cubes and then to the larger cubes. However, computer simulations [14] predict that $O_2$ molecules can be active and participate in nanoparticle surface growth (II) after $O_2$ molecules split over defects on the surface of growing MgO nanoparticles thus stimulating the most surface-defective incipient spherical shapes. The competition between the two growth mechanisms might take place by the intensification of the burning process that we performed.

It is known from literature that large terraced particles >200 nm can also grow in humid air by hydroxyl group passivation and thus stabilization of defective surfaces at later growth stage of primary cubes [15,16]. On contrary, we observe much smaller than 200 nm terraced nanoparticles and spheres down to 10 nm (Fig. 2). Such nanoparticles are also found for infrared laser intensified self-burning in air or oxygen where OH groups are not involved at all. This fact points out to mechanism (II) as the only reason for spherical and thus highly defective surface of nanoparticles in the environment of intensified self-burning. Then one may conclude that when spherical nanoparticles, initially grown on mechanism (II), are transported by convection to the environment favorable for MgO molecule condensation on mechanism (I) and continue their growth there, the incipient and then primary spheres start "healing" the surface defects and evolving into cubic nanoparticles through terrace step phase. It is easy to see that there is no other intermediate *evolutionary* shape between the initial sphere and the final cube except the *terraced* nanoparticle that preserves the cubic MgO lattice for the single-crystalline nanoparticle. However, in the case when growth mechanism (II) persists through the whole nanoparticle growth then larger perfect spheres can be grown from incipient and primary ones as we observe (Fig. 2).

From HR-TEM (one example is given in Fig. 1) we found a neat scaling law for terraced nanoparticles: the average size of steps is proportional to the size of the nanoparticle (Fig. 4). This behavior also excludes water vapor or hydrogen etching as the reason for the terraces as far as etching always leads to 1 nm steps as was mentioned above.

In order to understand the proportionality law let us consider the growth of terraced nanoparticles along the flame axis. Nanoparticles were thermophoretically sampled with the height step of 50 mm starting from the burner rim and the nanoparticle size $d_g$ was determined from TEM images (see Fig. 5). According to Ref. [16] for diffusion-limited surface growth, $d$ follows the law:

$d_g = [DV_m(C_b - C_d)t]^{1/2} + d^*$  (1)

where $D$ is the diffusion coefficient for MgO molecules that are nucleated in surface patches on growing nanoparticles; $V_m$ is the molar volume of MgO species; $C_b$ is the bulk concentration and $C_d$ is the concentration at equilibrium with nanoparticles of size $d_g$; $t$ is the time of surface growth and $d^*$ is the minimum size for a stable surface nucleus (patch) at $C_b$ assuming the diffusion path is much larger than the particle size and that MgO species follow free molecular transport. Here for simplicity we will consider only scaling regime and neglect any nucleus size for nanoparticles and later on for terrace steps. If one introduces apparent diffusion coefficient [5] $D^* = DV_m(C_b-C_d)$ and takes into account that the flame coordinate $x$ changes as $x=vt$ where $v=1$ m/sec is the gas velocity, one may rewrite Eq. (1) as

$$d_g=(D^*)^{1/2}t^{1/2}=(D^*/v)^{1/2}x^{1/2} \qquad (2)$$

The fit with Eq. (2) in Fig. 5 yields $D^*$=82.4 nm$^2$/ms to be compared with $D^*$=92.6 nm$^2$/ms for silica nanoparticles from Ref. [17]. Encouraged by this similarity we take for further estimations the value of $D$= 0.926 cm$^2$/sec at about 1900 K from Ref. [17].

Now let us consider the evolution of the terrace step size $d_s$. After a violent defect induced surface growth in pure oxygen leading to precursor spherical particles, the nanoparticle enters the flame region where absorption-desorption processes[17] dominate. The precursor spherical nanoparticle less than 5 nm being a pseudocubic nanoparticle with perfect MgO lattice according to HR-TEM may have as few as 10 elementary steps of the MgO lattice constant of 0.21 nm from each "side". These elementary steps (or grooves) act as incipient terrace steps (grooves) and may start developing into terrace steps (grooves) seen in Fig. 1. The number of steps (grooves) may conserve while the diffusion growth occurs. If this is the case, then the step (groove) size $d_s$ is automatically proportional to the nanoparticle size and the step growth follows Eq. (2) $d_s \sim t^{1/2}$. Meanwhile, the absorption-desorption theory [18] predicts the size of a right angle groove on the surface to be

$$d_s=1.13A^{1/2}t^{1/2} \qquad (3)$$

where $A=p_0\gamma V_m^{2}/[(2\pi m)^{1/2}(kT)^{3/2}]$; $p_0$ is the vapor pressure in equilibrium with the plane surface; $\gamma$ is the surface free energy per unit area; $m$ is the weight of the molecule; $kT$ is the thermal energy.

From Eq. (2) and Eq. (3) the proportionality law $d_s=1.13(A/D^*)^{1/2}d_g$ immediately follows. To compare it with the fit in Fig. 4 we have to find the proportionality constant in the law. In order to simplify the estimation of the constant we will approximate $(C_b-C_d)$ in Eq. (1) with the ideal gas value $p_0/kT$ and then $(A/D^*)^{1/2}=[\gamma V_m/(D(2\pi m\, kT)^{1/2})]^{1/2}$. By using the values [18] $\gamma\sim 10^3$ erg/cm$^2$; $m\sim 10^{-22}$g; $D\sim 1$cm$^2$/sec and $V_m\sim 10^{-22}$ cm$^3$ we get $d_s\sim 0.1 d_g$ to be compared with $d_s\sim 0.03 d_g$ from the fit in Fig. 4. Bearing in mind a large number of approximations made the comparison seems quite favorable for the suggested growth model for terraced MgO nanoparticles.

## 4. Conclusions

A morphology control of pure MgO nanoparticles by intensification of Mg burning has been presented. We showed that Mg fine powder burning in oxy-hydrogen flame or under the infrared laser beam generate mostly terraced and spherical MgO nanoparticles with the terrace steps scaling with the size of the nanoparticle. The growth mechanism for spherical and terraced nanoparticles has been described and compared with experiment.

## 5. Acknowledgements

This work was done by financial support from the Global Frontier Center for Multiscale Energy Systems supported by the Korean Ministry of Science and Technology (2011-0031561). We thank Daegyu Kim for assistance with the CO$_2$ laser experiment. Financial support from BK21 program and WCU(World Class University) multiscale mechanical design program(R31-2008-000-10083-0) through the Korea Research Foundation is gratefully acknowledged. The support from Institute of Advanced Machinery and Design, Seoul National University is also gratefully acknowledged.

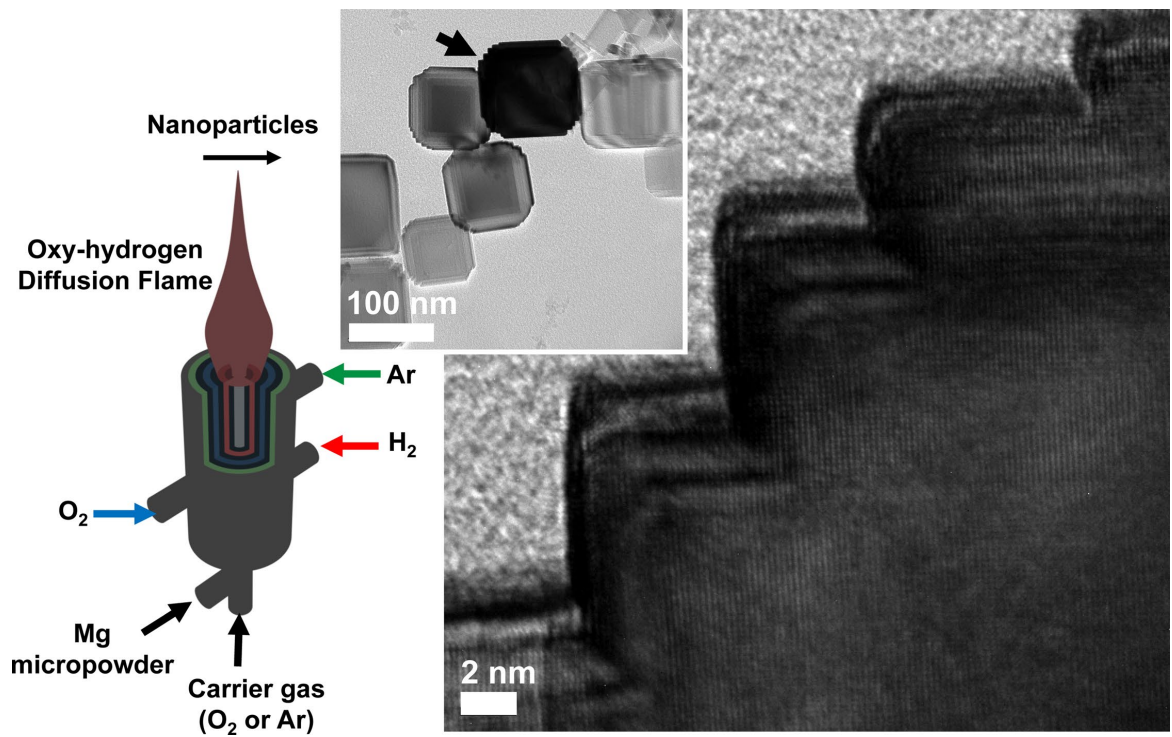

**Figure 1.** HR-TEM images of terraced nanoparticles grown in result of burning Mg micropowder in oxy-hydrogen diffusion flame shown in the sketch on the left. The black arrow in the central plate indicates the step train that was magnified.

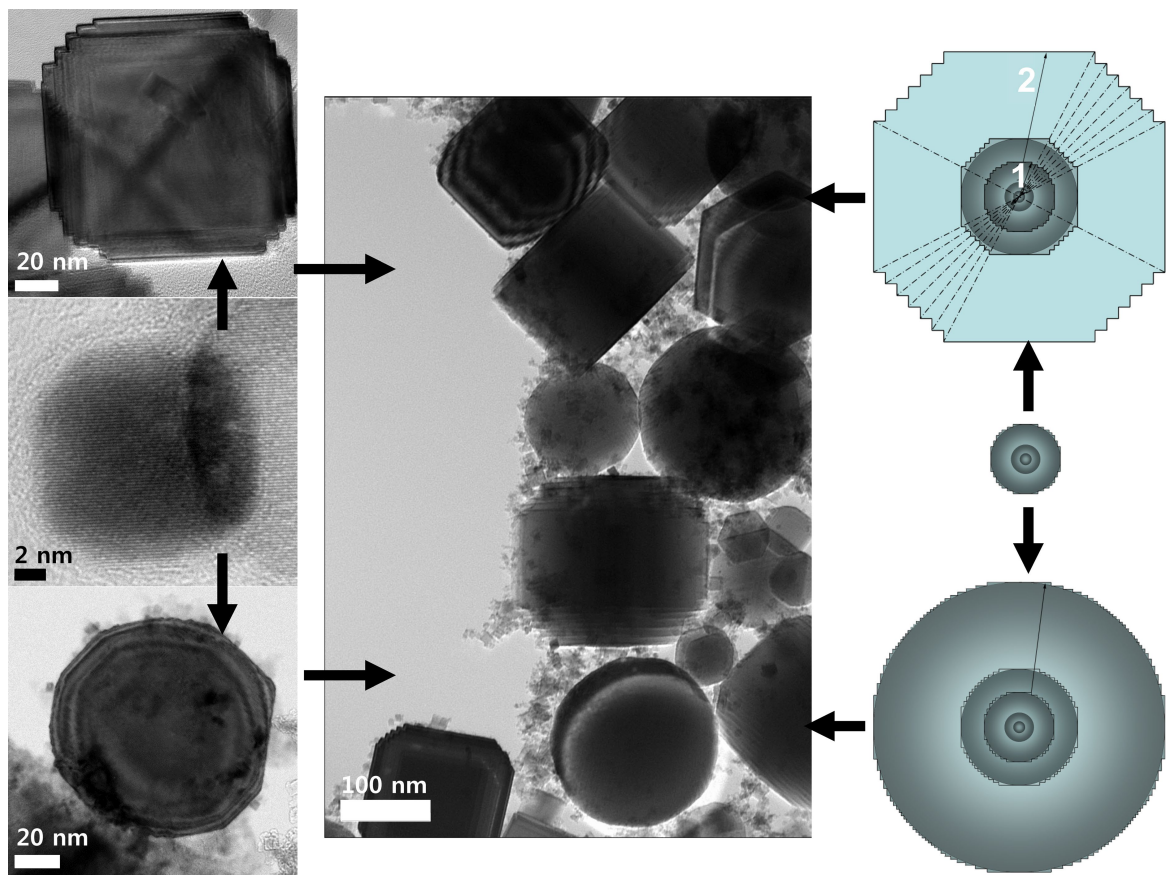

**Figure 2.** Illustration of the mechanism for generation of terraced MgO nanoparticles. The middle part of left (HR-TEM) and right (sketch) panels show a germ spherical nanoparticle created by fast surface defect induced growth mechanism (see text). The steps are of the size of the lattice distance. Top left and right panels show the spherical nanoparticle evolution (indicated with arrows) into a terraced nanoparticle in flame where adsorption-desorption of MgO molecules dominates. The step size increases proportionally to the particle size and the number of steps nearly conserves as indicated by dashed lines in the top right panel. The evolution of the nanoparticle starts from the germ nanoparticle marked 1 and continues to the large one marked 2. Bottom left and right panels show the spherical nanoparticle evolution into a much larger spherical particle when the germ nanoparticle continues its growth in $O_2$ environment where the surface defect induced growth dominates. The middle panel is the TEM image of the resulting MgO smoke where both large spheres and terraced nanoparticles are distinguished.

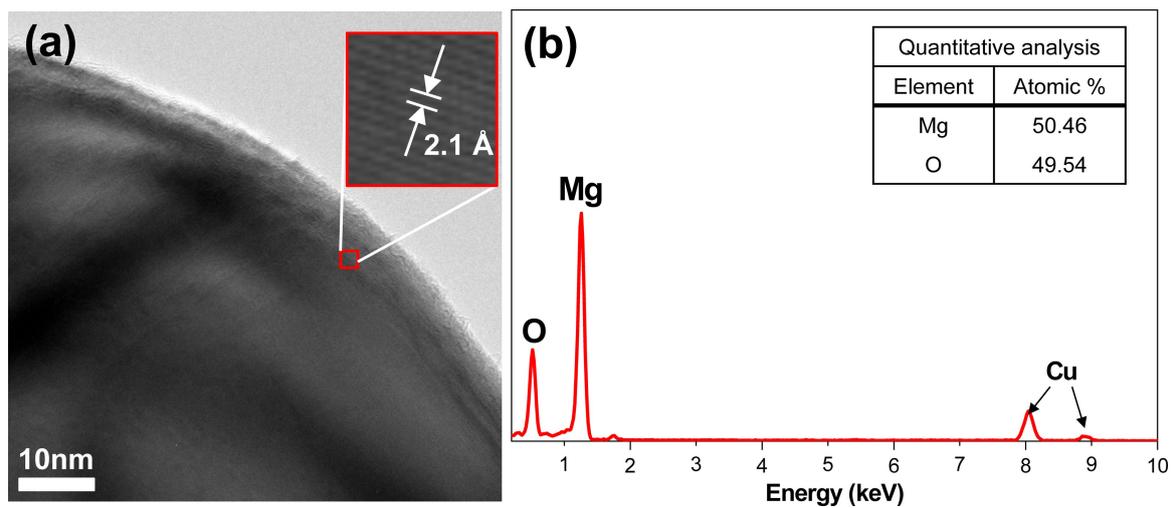

**Figure 3.** (a) An HR-TEM image of a spherical MgO nanoparticle. The distance 2.1 A is shown in the inset. (b) An EDX spectrum showing 1:1 correspondence of Mg to O ions.

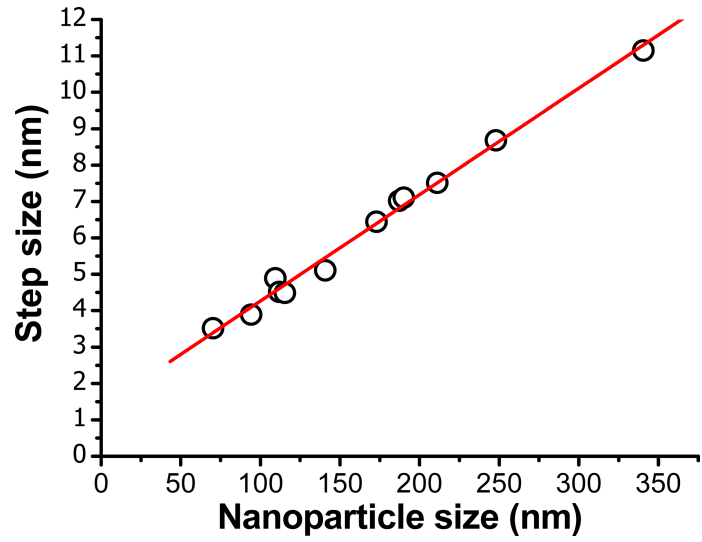

**Figure 4.** The step size $d_s$ averaged over each nanoparticle vs. the nanoparticle size $d_g$. The red line is the linear fit.

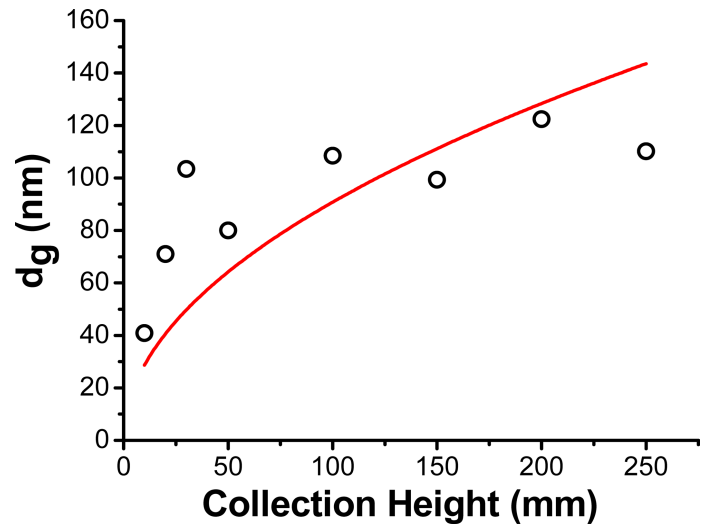

**Figure 5.** The nanoparticle size *d<sub>g</sub>* vs. the collection height *x* from the rim of the burner. The red line is the square-root fit according to Eq. (2) from the text.

Graphical Abstract

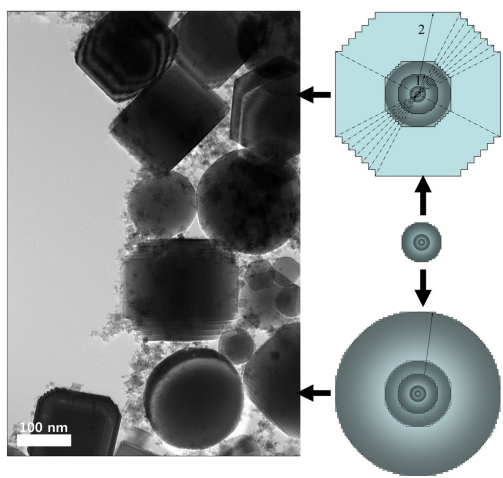